\documentclass[runningheads]{llncs}
\usepackage{orcidlink}
\usepackage[T1]{fontenc}
\usepackage{graphicx}
\usepackage{caption}
\usepackage{float}
\usepackage[inline]{enumitem}
\usepackage[colorinlistoftodos]{todonotes}

\usepackage{tabularx} 

\begin{document}
\title{Human--AI Collaboration for Scaling Agile Regression Testing: An Agentic-AI Teammate from Manual to Automated Testing}

\titlerunning{An Agentic-AI Teammate from Manual to Automated Testing}
%
\author{Moustapha El Outmani\inst{1}\orcidlink{0009-0003-1081-6038} \and
Manthan Venkataramana Shenoy\inst{1}\orcidlink{0009-0002-2756-1652} \and
Ahmad Hatahet\inst{1}\orcidlink{0009-0009-7677-7514} \and
Andreas Rausch\inst{1}\orcidlink{0000-0002-6850-6409} 
Tim Niklas Kniep\inst{2}  \and
Thomas Raddatz\inst{2}  \and
Benjamin King\inst{2}
}

\authorrunning{M. El Outmani et al.}
%

\institute{Institute for Software and Systems Engineering, Technical University of Clausthal, Adolph-Roemer-Straße 2A, 38678 Clausthal-Zellerfeld, Germany
\email{@tu-clausthal.de}\\
\url{https://www.isse.tu-clausthal.de/}
\and Hacon Ingenieurgesellschaft mbH – A Siemens Company, Lister Str. 15, 30163 Hannover, Germany
\email{info@Hacon.de}\\
\url{https://www.Hacon.de/}
}
\maketitle              

\begin{abstract}
Automated regression testing is essential for maintaining rapid, high-quality delivery in Agile and Scrum organizations. Many teams, including Hacon (a Siemens company), face a persistent gap: validated test specifications accumulate faster than they are automated, limiting regression coverage and increasing manual work. This paper reports an exploratory industrial case study of the Hacon Test Automation Copilot, an agentic AI system that generates system-level regression test scripts from validated specifications using retrieval-augmented generation and a multi-agent workflow. Integrated with Hacon’s CI pipelines, the Copilot operates asynchronously as a \textit{"silent AI teammate",} producing candidate scripts for human review. Mixed-method evaluation shows the AI accelerates script authoring and increases throughput, with 30–50\% code reuse. However, human review remains necessary for maintainability and correct domain interpretation. Clear specifications, explicit governance, and ongoing human–AI collaboration are critical. We conclude with lessons for scaling regression automation and enabling effective human–AI teaming in Agile settings.
\end{abstract}

\section{Introduction}
Agile software teams rely on continuous integration and automated regression testing to maintain quality while delivering features at speed \cite{10.5555/559553,forsgren2018accelerate,kim2018phoenix}. Yet, in large-scale environments such as Hacon (a Siemens company), the pace of evolving requirements and ongoing feature development outstrips the capacity to automate regression tests, resulting in a persistent gap between the volume of manual test specifications and the rate at which they are converted into executable scripts. This gap not only limits regression coverage and leads to increased manual testing effort in sprint releases, challenging the core agile principles of rapid feedback and sustainable delivery.

While recent advances in large language models (LLMs) and agentic AI offer promising capabilities for code and test generation \cite{augusto2025large}, there is limited industry guidance on integrating such tools into agile test engineering. Existing research often overlooks the nuanced requirements of regression test suites and rarely addresses the human factors of collaboration, review, and governance in real-world teams \cite{neumann2026between,tawosi2025almas,martinfowlerUnderstandingSpecDrivenDevelopment,qodoSoftwareTesting}.

This paper presents an exploratory industrial case study with Hacon's agile test engineering team, examining the introduction of an agentic AI system for automated test script generation and its integration into the team's workflow. We address the following research question:

\begin{enumerate}
    \item[\textbf{\textit{RQ:}}] How does integrating an agentic AI system for automated test generation affect productivity, test script quality, and human–AI collaboration within an agile testing team?
\end{enumerate}

We introduce an agentic AI teammate that generates executable system test scripts from validated specifications using retrieval-augmented generation (RAG) and a bounded multi-agent workflow, integrated with Hacon’s CI pipelines and supporting human review and artifact governance. Evaluation is based on artifact analysis and practitioner feedback.

Our contributions are:
(\textit{C1}) We present the design of an agentic AI system that transforms validated manual test specifications into executable regression test scripts within an agile testing workflow.
(\textit{C2}) We describe governance, collaboration, and review practices
for integrating an AI teammate into agile test engineering workflows.
(\textit{C3}) We empirically evaluate the impact of this approach on
productivity, test script quality, and human--AI collaboration through
an industrial case study.

The rest of the paper is structured as follows: 
Section~\ref{sec:industry-context} describes the industrial context and the manual-to-automated testing bottleneck at Hacon. 
Section~\ref{sec:sys-design} presents the architecture of the Hacon Test Automation Copilot. 
Section~\ref{sec:coll-pattern} explains how the Copilot is integrated into the agile testing workflow and how human–AI collaboration is organized. 
Section~\ref{sec:eval-results} reports the empirical evaluation and key findings from the industrial case study. 
Section~\ref{sec:discussion} discusses implications, limitations, and directions for future work.

\section{Industrial Context and Problem Statement} \label{sec:industry-context} 

Large industrial software organizations developing complex, long-lived systems increasingly rely on automated regression testing to manage scale and frequent change \cite{regressionTest1997,regressionTest2012,agileMethodologies2012}. Hacon (a Siemens company) represents such an industrial setting, developing large-scale transportation planning and scheduling software within a mature agile development organization, delivering frequent releases through short development sprints supported by robust CI pipelines and established system-level test automation. In this environment, regression testing involves repeatedly executing system-level tests to ensure that newly introduced changes do not break existing functionality. These tests are continuously integrated and executed within the shared CI pipeline, with the studied test engineering team collaborating closely with development teams throughout the agile development process.
 
Within Hacon’s testing workflow, test specifications are first created and validated by test engineers before being manually translated into executable automation scripts. This translation step requires significant engineering effort and domain expertise, resulting in a bottleneck where the number of validated test specifications grows faster than the team’s capacity to automate them.

\begin{figure*}[htbp]
    \centering
    \includegraphics[width=\textwidth]{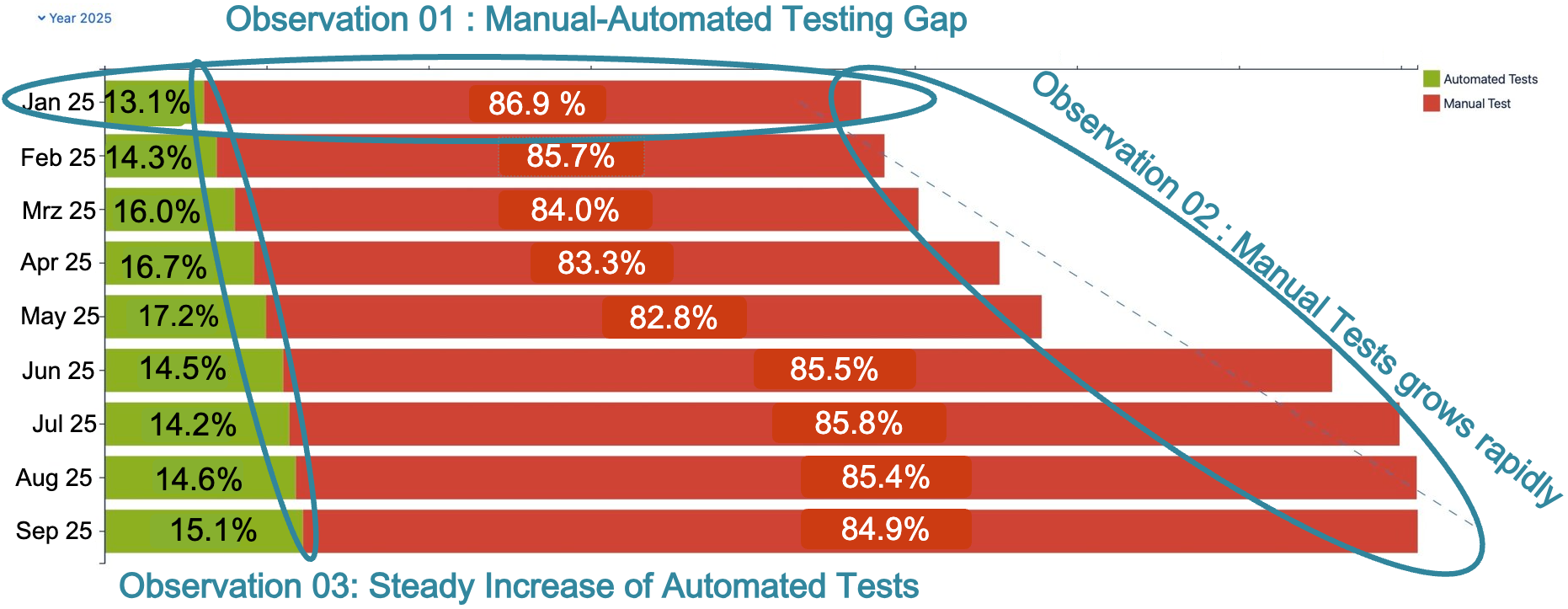}
    \caption{Growth of manual and automated system tests across monthly releases in 2025. Manual tests dominate the regression suite, while automated coverage increases gradually, illustrating the persistent gap between manual specifications and automated scripts.}
    \label{fig1}
\end{figure*}

Despite continuous investment in automation and dedicated test engineering teams, Hacon faces a persistent challenge: manual test specifications continue to outpace the rate at which they are automated. As shown in Figure~\ref{fig1}, automated test coverage increases steadily by 1–2\% per release, yet manual tests still account for 82–87\% of the total, with their absolute number growing by 10–20\% per release. This ongoing automation gap results in increased manual testing effort in the regression suite and slower feedback cycles, motivating the need for AI-assisted approaches that can accelerate script creation while upholding Hacon’s high standards for quality, maintainability, and governance.

\section{System Architecture: Hacon Test Automation Copilot} \label{sec:sys-design}

To address the automation bottleneck described in Section~\ref{sec:industry-context}, we developed the \textit{Hacon Test Automation Copilot}, an agentic AI system that generates executable system-level test scripts from validated specifications. The Copilot operates as an asynchronous, batch-oriented collaborator, preparing candidate automation artifacts ahead of sprint activities and integrating with Hacon’s continuous integration (CI) workflow.

\begin{figure*}[htbp]
    \centering
    \includegraphics[width=\textwidth]{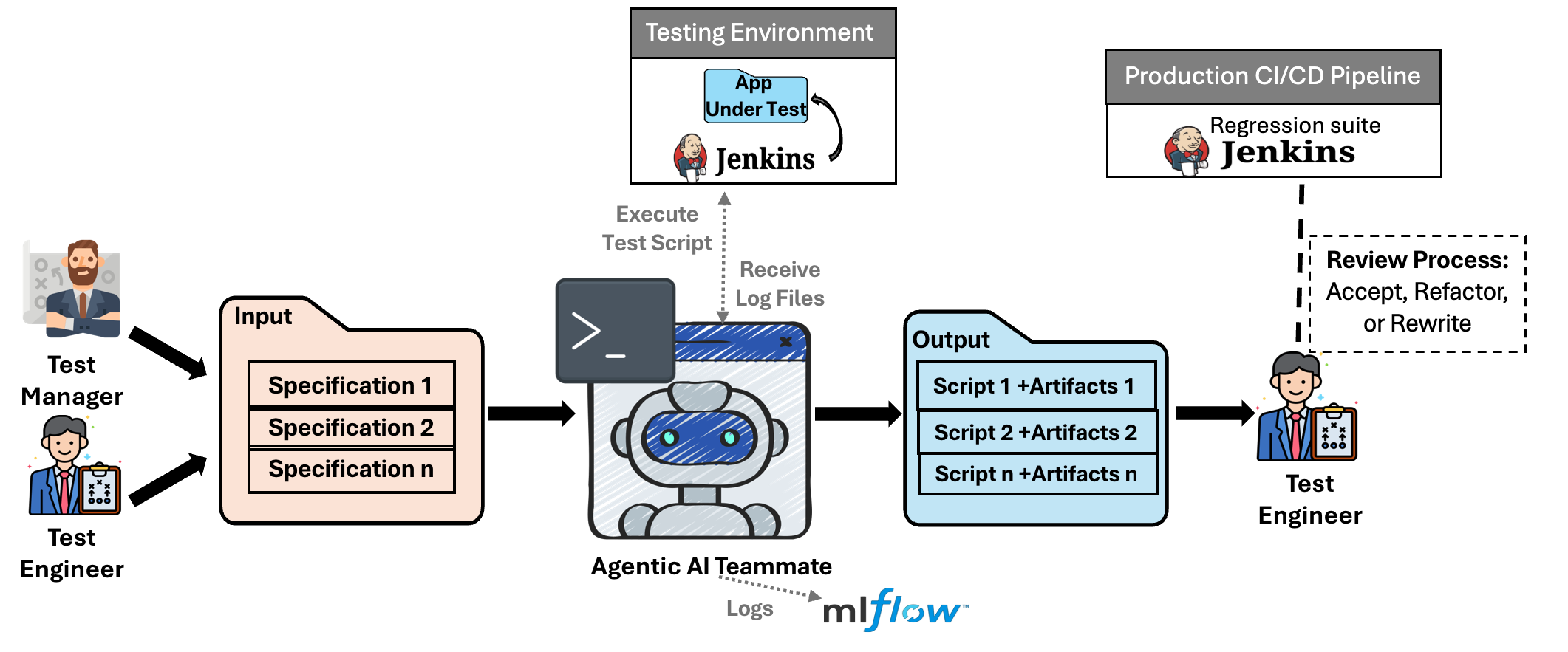}
     \caption{Overview of the Hacon Test Automation Copilot architecture and its interaction with test engineers. The system generates candidate test scripts from validated specifications, executes them in the CI environment, evaluates the results, and produces review-ready artifacts for human validation before integration into the regression suite.}
    \label{fig2}
\end{figure*}

Test specifications are placed in a designated input folder and processed via a command-line trigger. As shown in Figure~\ref{fig2}, the Copilot processes each specification and produces review-ready artifacts: the generated script, Jenkins execution logs, structured summaries, and execution traces in MLflow. Test engineers review these outputs to decide on acceptance, refactoring, or rewriting before adding scripts to the regression suite.

The Copilot’s architecture implements a bounded generation–execution–evaluation loop, orchestrated to respect iteration limits and CI resource availability. Within each cycle, a \textbf{Generator agent} uses retrieval-augmented generation (RAG) over historical specification–script pairs to create candidate test scripts. These are executed in a Jenkins environment, with the application under test and logs collected. An \textbf{Evaluator agent} analyzes execution logs for syntactical and semantic correctness, coverage, and improvement potential. A \textbf{Reporter agent} consolidates results into structured summaries and execution reports for both test managers and engineers. All traces are recorded in MLflow for traceability.

By operating asynchronously and \textbf{autonomously}, the agentic AI teammate generates candidate automation artifacts in advance, allowing engineers to review them without interrupting ongoing development activities. The system is configurable and supports maintainability by expanding the RAG base and updating prompts. The architecture is designed for seamless integration and future extensibility.

\section{Integration and Human--AI Collaboration in the Agile Testing Workflow}
\label{sec:coll-pattern}

The Copilot was integrated into Hacon’s agile testing workflow as a \textit{silent AI teammate}, operating asynchronously to generate candidate automation scripts before each sprint. This fits the team’s sprint-based process, where regression tests must be ready and validated quickly. Instead of continuous interaction, the Copilot produces first-draft artifacts for later review and refinement by test engineers prior to inclusion in the regression suite.

Test specifications, created and validated by engineers and managers, are exported as input to the Copilot. The AI system generates candidate scripts, which the test automation team reviews at the sprint’s start. This setup lets the AI handle repetitive script generation and initial validation, while engineers focus on interpretation, refinement, and approval. Engineers review artifacts and execution evidence to ensure tests match the intended behavior. Scripts are modified or rewritten as needed before acceptance.

Governance is ensured by explicit human oversight and artifact traceability. The Copilot has \textit{bounded autonomy}: it cannot add scripts to the regression suite without human approval, and all outputs are recorded. This review process keeps automated test generation aligned with quality standards and maintains accountability.

\section{Evaluation and Results} \label{sec:eval-results}

This section presents the empirical evaluation of the agentic AI system’s impact on productivity, test script quality, and human–AI collaboration in Hacon’s agile test engineering workflow.

\subsection*{Study Design and Data Collection}

The evaluation was conducted over a four-week period with Hacon’s test engineering team (five engineers). We selected 61 test specifications (Xray format), spanning six functional areas and varying in complexity (2–18 steps per case, story point estimates 3–8, and pre-rated input clarity from A to D). For each, the Copilot generated an initial test script, which was then reviewed and refactored by a human engineer. Data collection included:
\begin{itemize}
    \item \textbf{Structured Survey:} Engineers rated each script using a standardized form, assessing block-level correctness, overall quality, readability, usability, and test smells (1–5 Likert scale), and provided open-text recommendations. The survey also covered the clarity, completeness, accuracy, and actionability of the generated markdown reports.
    \item \textbf{Artifact Comparison:} For a focused subset of approximately 20 representative cases, two expert test engineers conducted a detailed artifact comparison. Using version control, they systematically compared the AI-generated scripts to the final, human-corrected versions, assessing semantic equivalence, maintainability, and recurring issues.
\end{itemize}

\subsection*{Key Results and Insights}
\paragraph{1. Productivity}
Artifact comparison revealed that, on average, 30–50\% of the AI-generated code per script was retained unchanged by test engineers. This indicates that the AI teammate can provide a substantial head start in script authoring, reducing the time required to move from specification to executable test.

\paragraph{2. Test Script Quality}
 Most AI-generated scripts needed moderate to major human edits: of 49 scripts, 15 were fully rewritten, 20 required major, 13 medium, and 1 minor changes. Common issues included: ‘...test data should be externalized...’, ‘...unnecessary imports...’. Typical issues were hardcoded data (31/49), redundant imports (29/49), unused objects (23/49), and missing or incorrect validations (over 30 cases). These issues primarily affected maintainability, readability, and long-term usability rather than basic functional correctness. Such quality concerns are critical in Agile and Scrum environments, where regression suites must remain stable, maintainable, and easy to evolve alongside frequent releases.

\paragraph{3. Human–AI Collaboration}
Version control analysis of all corrected scripts, together with qualitative feedback, highlighted a dynamic collaboration pattern. Engineers frequently replaced AI-generated code blocks with established, trusted snippets from previous projects, even when the AI’s logic was technically correct. This behavior reflected a preference for familiar and maintainable code. The artifact comparison further showed that the AI’s literal interpretation of specifications—stemming from the fact that test specs were designed for humans who intuitively know what to ignore or adapt. Human review remained essential, especially to address implicit domain knowledge and ensure production readiness.

\section{Discussion, Lessons Learned \&
Future work} \label{sec:discussion}

\paragraph{\textbf{Discussion.}}
This exploratory industrial case study shows that integrating an agentic AI teammate into an Agile regression testing workflow can accelerate the conversion of validated test specifications into automated scripts, but it also exposes significant limitations. While the AI system provided a measurable productivity boost—helpful in about 30\% of evaluated cases—most generated scripts required moderate to major human intervention before they could be used in the regression suite. Client feedback highlighted that the agent often produced solutions that were “technically correct, which in this case is a bad kind of correct,” meaning the scripts met the literal requirements but failed to align with maintainability and domain-specific expectations valued by Agile and XP teams. The client also questioned the representativeness of the test set, the completeness of test descriptions, and identified the need for further evaluation and possibly explicit Framework domain convention support. These points challenge the generalizability of our findings and suggest that the effectiveness of AI-driven automation is highly dependent on the quality and clarity of input specifications, as well as on ongoing human review.

The results also surfaced key assumptions that do not always hold in real-world Scrum and Agile environments. The AI system’s performance relied on the premise that test specs are complete, data is fully known and available, and existing automation scripts are both accurate and stable—assumptions that practitioners recognized as fragile. The agent’s literal approach exposed ambiguities and gaps that human testers routinely resolve using tacit knowledge, reinforcing that human oversight remains indispensable for ensuring semantic correctness, maintainability, and trust in the regression suite. While the client’s goal is to close the automation gap with lower cost and moderate investment by “automating the automation with AI,” our findings indicate that sustainable gains require not only technical advances but also improved specification practices and explicit human–AI collaboration. Future work should focus on co-adaptation: enhancing AI capabilities through feedback and better prompts, and supporting teams in authoring clearer, more actionable test specifications.

\paragraph{\textbf{Lessons Learned.}}
key lessons emerged from deployment and evaluation:

\begin{itemize}
    \item \textbf{Specification Clarity is Foundational:} AI effectiveness depends on clear, complete specifications. Human engineers routinely bridge gaps using tacit knowledge, but the AI requires unambiguous, comprehensive instructions. 
    
    \item \textbf{AI Follows, Humans Lead:} The AI’s literal interpretation of specifications can lead to “technically correct” but contextually inappropriate solutions, underscoring the need for domain-informed review.

    \item \textbf{Human Review Is Indispensable:} Despite automation, human oversight is indispensable for ensuring semantic correctness, maintainability, and trust in the regression suite.
    
    \item \textbf{Assumptions Must Be Challenged:} The project surfaced several implicit assumptions—about test set representativeness, data availability, and maintenance effort—that require ongoing scrutiny as the system evolves.
    
    \item \textbf{Co-Adaptation is Key:} Sustainable productivity gains will require both improved AI adaptation (via better prompts, feedback loops, and semantic evaluation) and changes in human practices (such as specification authoring and review routines).
    
\end{itemize}

\paragraph{\textbf{Limitations.}}
The study is limited to one company, a small team, and a specific set of scripts. Results may not generalize to other settings. Practitioner feedback may be biased, and the evaluation period was short. Technical choices in orchestration, prompt engineering, and integration may affect reproducibility.

\paragraph{\textbf{Future Work.}}
Future work will focus on co-adaptation between humans and AI. We plan to (1) improve AI adaptation by integrating expert feedback and adding specialized agents for semantic evaluation, and (2) support teams in standardizing specification practices and authoring clearer test steps. These steps aim to further reduce the gap between specifications and scripts, and to strengthen human–AI collaboration for more effective and scalable regression automation.

\begin{credits}

\subsubsection{\ackname}
We gratefully acknowledge the partnership and support of Hacon (a Siemens company) in providing the industrial context, data, and invaluable practitioner feedback for this study. Their collaborative engagement was essential to the system’s deployment and evaluation.
\end{credits}

\begin{credits}

\subsubsection{\discintname}
All authors declare that they have no conflicts of interest.
 
\end{credits}


\bibliographystyle{splncs04}
\bibliography{mybibliography}

\end{document}